\pdfoutput=1
\documentclass[conference,letterpaper]{IEEEtran}
\IEEEoverridecommandlockouts%
\usepackage[T1]{fontenc}%
\usepackage{amsmath,amssymb}
\usepackage{graphicx}
\usepackage{booktabs,multirow}
\usepackage{stfloats}
\usepackage{xcolor}
\usepackage{tikz}
\usepackage[american]{circuitikz}%
\usetikzlibrary{arrows.meta,positioning,fit,calc,backgrounds}
\usepackage{listings}
\usepackage{cite}
\usepackage{url}

\usetikzlibrary{arrows.meta,positioning,fit,calc,backgrounds,shapes.geometric,shapes.multipart,decorations.pathreplacing,matrix}

\begin{document}

\title{AnalogScout: Simulation-Efficient PVT-Aware\\Analog Sizing with LLM Agents}

\author{\IEEEauthorblockN{Yichen Long}
\IEEEauthorblockA{\textit{College of Integrated Circuits and Micro-Nano Electronics}\\
\textit{Fudan University}\\
Shanghai, China\\
longyc26@m.fudan.edu.cn}
\and
\IEEEauthorblockN{Keren Zhu\textsuperscript{*}}
\IEEEauthorblockA{\textit{College of Integrated Circuits and Micro-Nano Electronics}\\
\textit{Fudan University}\\
Shanghai, China\\
krzhu@fudan.edu.cn}
\thanks{\textsuperscript{*}Corresponding author: Keren Zhu (krzhu@fudan.edu.cn). This work has been submitted to the IEEE for possible publication. Copyright may be transferred without notice, after which this version may no longer be accessible.}}

\maketitle

\begin{abstract}
Analog device sizing must satisfy all specifications across process, voltage, and temperature (PVT) corners to ensure tape-out success, but simulating every candidate at all corners dominates the sizing time and slows design iteration.
Conventional surrogate-based optimizers search the sizing space directly, while recent large language model (LLM) agents add circuit knowledge to designs but still delegate the sizes to global optimizers.
Both learn the circuit response from repeated multi-corner SPICE simulations and optimize on top of it, which becomes expensive under a limited simulation budget.
Providing an agent with reusable verified designs and sizing sub-blocks analytically could save most of this budget.
We propose AnalogScout, a simulation-efficient PVT-aware sizing framework driven by an LLM agent.
Combining elite-design reuse, analytic gm/Id local sizing and global refinement under full-corner verification, AnalogScout reduces power by 43.3\% relative to the evolution strategy CMA-ES at an equal budget of 16k simulations on three SKY130 OTAs of increasing complexity, and reaches the final CMA-ES power after 7.0\% of that budget.
It also attains lower power than Bayesian optimization on all three OTAs and than two re-implemented LLM-driven flows on the two OTAs where they were tested.
\end{abstract}


\section{Introduction}
Analog design automation lags far behind its digital counterpart, and device sizing remains its most labor-intensive step~\cite{gielen_cad}.
Sizing selects the widths, lengths and bias currents that meet gain, bandwidth, stability and noise targets at minimum power.
In advanced nodes these targets must hold at every process, voltage and temperature (PVT) corner, so each candidate costs a full multi-corner SPICE sweep.

\begin{figure}[t]
\centering
\includegraphics[width=\columnwidth]{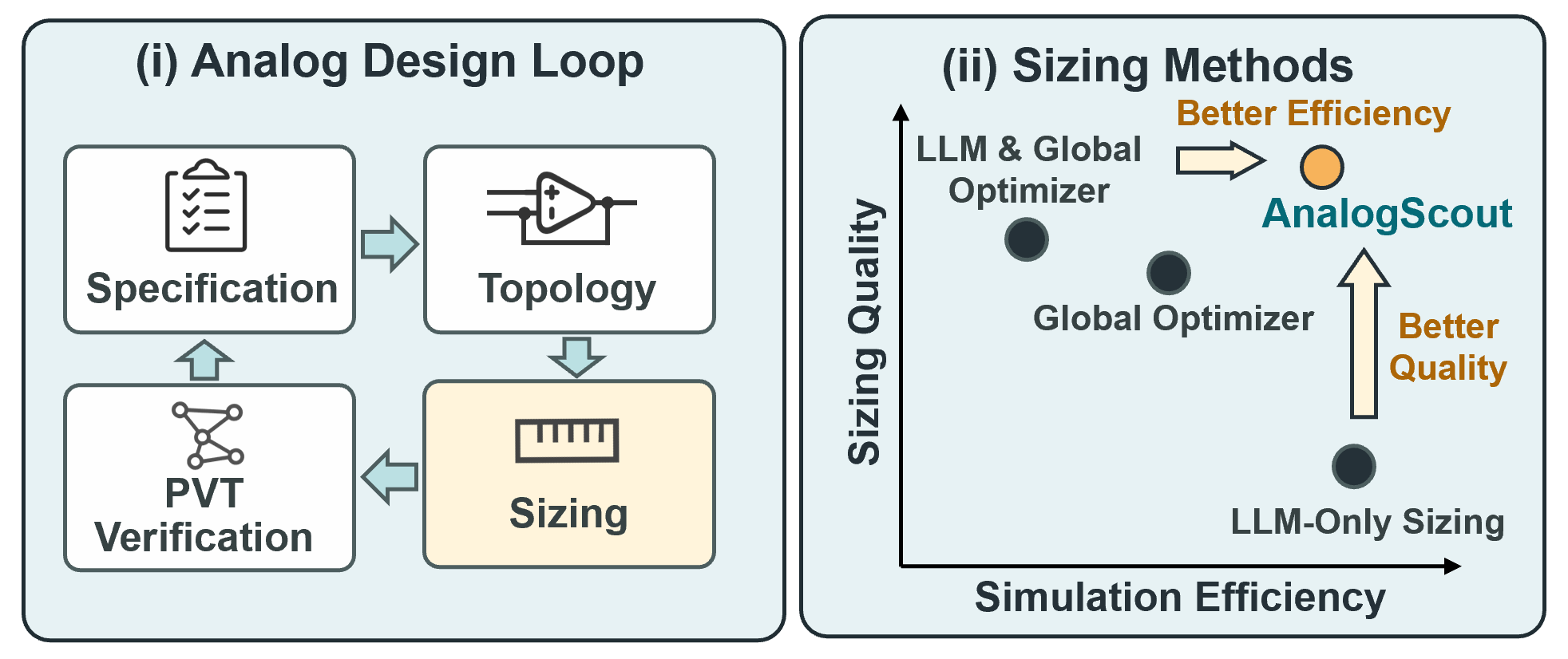}
\caption{(i) In the analog design loop, a topology chosen for the specification is sized and verified at all PVT corners until every specification holds; the sizing step (highlighted) consumes most simulations. (ii) Qualitative positions of sizing methods: LLM-only sizing consumes fewer simulations; a global optimizer, alone or driven by an LLM, sizes well with many simulations. AnalogScout targets both quality and efficiency.}
\label{fig:motivation}
\end{figure}

\begin{figure*}[t]
\centering
\begingroup%
\colorlet{asLLM}{blue!10}%
\colorlet{asTool}{yellow!25}%
\colorlet{asSim}{gray!15}%
\begin{tikzpicture}[
  font=\fontsize{7.2}{8.3}\selectfont,
  >={Stealth[length=3pt,width=2.6pt]},
  box/.style={draw=black!75,line width=.4pt,rounded corners=1.5pt,
    align=center,inner xsep=3pt,inner ysep=2pt},
  col/.style={box,minimum width=3.45cm},
  ev/.style={box,fill=asSim,minimum width=4.0cm,minimum height=.36cm},
  phase/.style={box,fill=asTool,minimum width=7.0cm,minimum height=1.24cm},
  arr/.style={->,line width=.45pt,draw=black!80},
  lab/.style={text=black!85,inner sep=1pt,align=center},
  frame/.style={draw=black!55,dashed,line width=.4pt,rounded corners=3pt},
  ttl/.style={fill=white,inner sep=1pt,anchor=west}
]
\def\asKey#1{{\setlength{\fboxsep}{0pt}\setlength{\fboxrule}{.4pt}%
  \raisebox{-.3pt}{\fcolorbox{black!65}{#1}{\rule{0pt}{5.5pt}\rule{5.5pt}{0pt}}}}}

\node[ev,fill=asTool,anchor=north] (cache) at (7.0,-.35)
  {\textbf{Evaluation Cache}\\Exact-Match Reuse};
\node[ev,fill=asTool,below=.10cm of cache] (c1)
  {\textbf{Compilation}\\Isolated Execution};
\node[ev,fill=asTool,below=.10cm of c1] (c2)
  {\textbf{gm/Id Local Sizing}\\Analytic Solve (Phase-Localize)};
\node[ev,fill=asTool,below=.10cm of c2] (c3) {Design Rule Check};
\node[ev,below=.10cm of c3] (c4) {Bias / Start-Up Check};
\node[ev,below=.10cm of c4] (c5) {Saturation Check};
\node[ev,below=.10cm of c5] (c6) {Nominal Spec Check};
\node[ev,below=.10cm of c6] (c7) {Full-PVT Check};
\foreach \ca/\cb in {cache/c1,c1/c2,c2/c3,c3/c4,c4/c5,c5/c6,c6/c7}
  \draw[arr] (\ca) -- (\cb);

\node[col,fill=asLLM,anchor=north] (dsl) at (1.80,-.35)
  {\textbf{Design Source}\\Topology and Block Choices\\gm/Id Sizing Intent};
\node[col,fill=asTool,anchor=south] (pool) at (1.80,0 |- c7.south)
  {\textbf{Elite Pool}\\Elite-Biased Run Starts\\Lessons};
\node[col,fill=asLLM] (agent) at ($(dsl.south)!.5!(pool.north)$)
  {\textbf{LLM Agent}\\Edit and Evaluate\\Device Data and Building Blocks};
\begin{scope}[on background layer]
  \node[frame,solid,draw=black!70,fill=white,fit=(cache)(c7),
    inner xsep=.12cm,inner ysep=.15cm] (ev) {};
  \node[frame,fit=(dsl)(agent)(pool)(ev),
    inner xsep=.12cm,inner ysep=.14cm] (loop) {};
\end{scope}
\node[ttl] at ([xshift=4pt]loop.north west) {\textbf{Phase-Explore / Phase-Localize:} Agent Loop};
\node[fill=white,inner sep=1pt,font=\fontsize{7.2}{8.3}\selectfont\bfseries] at (ev.north)
  {Fixed Evaluator};
\draw[arr] (agent) -- node[lab,right] {Edits} (dsl);
\draw[arr] (pool) -- node[lab,right] {Elites and Lessons} (agent);
\draw[arr] (dsl.east |- cache) -- (cache.west -| ev.west);
\draw[arr,densely dotted] (dsl.south east) -- ++(.16cm,0)
  |- (c2.west -| ev.west);
\node[lab,above=1pt] at ($(c2.west -| ev.west)+(-.65cm,0)$)
  {Sizing\\Intent};
\draw[arr] ([yshift=2.5pt]agent.east)
  -- node[lab,above] {Evaluate} ([yshift=2.5pt]agent.east -| ev.west);
\draw[arr] ([yshift=-2.5pt]agent.east -| ev.west)
  -- node[lab,below] {Diagnostics} ([yshift=-2.5pt]agent.east);
\draw[arr] (c7.west -| ev.west)
  -- node[lab,above] {Verified\\Designs} (c7.west -| pool.east);

\node[phase,anchor=north west] (sweep) at (10.65,-.45)
  {\textbf{Phase-Restructure:} Structure Sweep\\
   Recognize Mirrors / Pairs / Compensation\\
   Swap Within Family\\Local Re-Size};
\node[phase,below=.30cm of sweep] (refine)
  {\textbf{Phase-Refine:} Global Refinement\\
   Matched-Group Search Around Best Design\\Restart on Stagnation};
\node[phase,fill=asSim,below=.30cm of refine] (signoff)
  {\textbf{Sign-Off}\\TT Mismatch Monte Carlo\\
   All-Corner Saturation\\Independent Re-Measurement};
\draw[arr] (sweep) -- (refine);
\draw[arr] (refine) -- (signoff);
\coordinate (detTop) at ([yshift=-.2pt]sweep.north |- loop.north);
\coordinate (detBot) at ([yshift=.2pt]signoff.south |- loop.south);
\begin{scope}[on background layer]
  \node[frame,solid,fit=(sweep)(detTop)(detBot),
    inner xsep=.12cm,inner ysep=0pt] (det) {};
\end{scope}
\node[ttl] at ([xshift=4pt]det.north west) {\textbf{Deterministic Flow}};
\draw[->,line width=1.3pt,draw=black!80,>={Stealth[length=6pt,width=6pt]}]
  (loop.east |- sweep.west) -- node[lab,above=1pt] {Top\\Designs} (det.west |- sweep.west);
\node[fill=white,inner xsep=1pt,inner ysep=0pt,anchor=east] at ([xshift=-4pt]det.north east)
  {\asKey{asLLM}\,Agent\quad\asKey{asTool}\,Tool\quad\asKey{asSim}\,Simulation};
\end{tikzpicture}%
\endgroup%

\caption{AnalogScout workflow. The LLM agent edits the design source and states gm/Id sizing intent; the fixed evaluator reuses cached results, resolves the intent analytically and checks each design up to full PVT; the top designs then pass through Phase-Restructure, Phase-Refine and sign-off.}
\label{fig:pipeline}
\end{figure*}
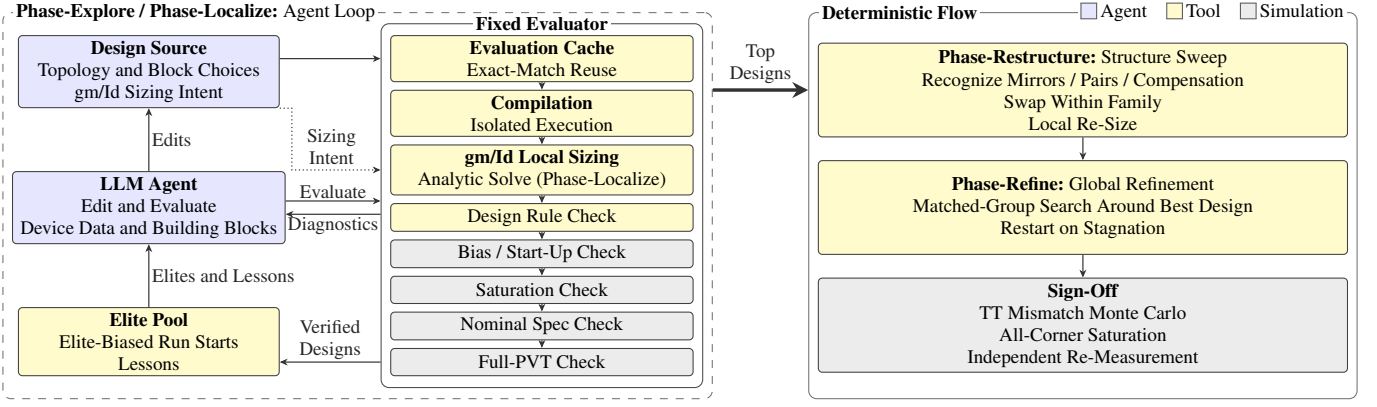

PVT-aware sizing searches for device sizes that satisfy all specifications at all corners, mainly through surrogate-based and population-based optimization.
Surrogate methods such as Bayesian optimization (BO) fit the corner-dependent response~\cite{wang_bo,lyu_mace,guo_bnnbyo}, whereas evolutionary and reinforcement-learning methods~\cite{liu_gaspad,autockt,gcnrl} search the design space corner by corner~\cite{li_pvttransfer,shi_robustanalog}.
Both learn the response from sampled simulations, so their cost scales with corners times iterations (Sec.~\ref{sec:prelim}).

LLM agents~\cite{analogcoder,artisan} add circuit knowledge to this search but spend the simulation budget poorly (Fig.~\ref{fig:motivation}).
LLM-only sizing loops oscillate between solutions~\cite{liu_newcas}, and most agentic flows delegate every device size to a global optimizer~\cite{analogcoderpro,adollm,eesizer,llmampsizing}.
We further observe that agents re-submit identical designs, restart instead of refining verified ones, and lose most of the budget to simulation-driven local optimizers.
An agent that reuses verified designs and sizes sub-blocks analytically would instead spend its budget on improving designs.

Inspired by agentic exploration in digital RTL design~\cite{rtlscout}, we propose AnalogScout, a simulation-efficient PVT-aware sizing framework, with three contributions:
\begin{itemize}
\item \textbf{Budget-aware exploration:} an elite pool and an exact evaluation cache let the agent refine verified designs instead of repeating or restarting them.
\item \textbf{Low-simulation local sizing:} the agent states gm/Id goals for sub-blocks, which a lookup-table solver resolves with at most three operating-point simulations per block.
\item \textbf{Results:} on three SKY130 OTAs verified at nine corners, AnalogScout reduces power by 43.3\% relative to CMA-ES at an equal 16k-simulation budget, matches the final CMA-ES power after 7.0\% of that budget, and on the Miller and NMC OTAs beats CMA-ES with twice the budget.
\end{itemize}
Section~\ref{sec:prelim} formulates PVT-aware sizing and reviews existing optimizers, Section~\ref{sec:method} details the four phases of AnalogScout, Section~\ref{sec:exp} compares it with optimizer and LLM baselines and traces where the budget goes, and Section~\ref{sec:concl} concludes.

\newpage%
\section{Preliminaries}\label{sec:prelim}
\textbf{PVT-aware sizing.}
Given a topology with device sizes $\mathbf{x}$, PVT-aware sizing solves
\begin{equation}
\min_{\mathbf{x}}\ P_{\mathrm{tt}}(\mathbf{x})\quad\text{s.t.}\quad f_i(\mathbf{x},c)\ \text{meets}\ s_i\quad\forall i,\ \forall c\in\mathcal{C},
\end{equation}
where $P_{\mathrm{tt}}$ is the power at the typical corner, $f_i$ are measured performances such as gain and phase margin, $s_i$ their specifications, and $\mathcal{C}$ the set of process, voltage and temperature corners.
Each candidate is therefore evaluated by $|\mathcal{C}|$ sets of SPICE simulations, and we measure every method by the number of simulations it consumes.

\textbf{PVT-aware optimizers.}
Surrogate-based methods fit a model of the corner-dependent response and optimize it, for example a Bayesian neural network shared across corners~\cite{guo_bnnbyo} or a contextual Gaussian process over design and PVT variables~\cite{huang_croza}.
Population-based methods evolve designs per corner and share knowledge across corners~\cite{li_pvttransfer}, and reinforcement-learning methods train one policy over all corners~\cite{shi_robustanalog,kong_pvtsizing}.
To reduce the corners simulated, these methods simulate only the current worst corners~\cite{pascu_smartcorner} or cluster similar corners~\cite{shi_robustanalog,kong_pvtsizing}.

\section{AnalogScout}\label{sec:method}
AnalogScout is a simulation-efficient PVT-aware sizing framework in which an LLM agent decides where and toward which goal to size, while deterministic tools compute the sizes and verify the design (Fig.~\ref{fig:pipeline}).
Its input is a topology with starting sizes, the specifications, a PDK and a simulation budget; it seeks a netlist that meets every specification at every PVT corner with low typical-corner power.
In Phase-Explore the agent rewrites the netlist, starting from the best verified designs so far; in Phase-Localize it attaches gm/Id goals to sub-blocks, which a lookup-table solver resolves.
Phase-Restructure then swaps standard sub-circuits within their family, and Phase-Refine spends the remaining budget on a global search around the best design before sign-off.

\subsection{Design Representation and Verification}
A design is a Python netlist built from PDK MOSFETs, bounded R and C, verified building blocks and precomputed gm/Id tables~\cite{silveira_gmid,gmid,murmann_iscas24}.
A fixed evaluator, which the agent cannot modify, solves the DC operating point and checks saturation and the specifications at the typical corner; a design that passes is verified at nine PVT corners (tt; ss, sf and fs at $-10\%$ and ff at $+10\%\,V_{DD}$, each at $-40/125^\circ$C) with fixed testbenches for every specification of Fig.~\ref{fig:benchmarks}.
A design is \emph{feasible} if every specification holds at every corner.
Its violation is
\begin{equation}
V=\sum_i \max_{c}\max\!\left(0,\,g_{i,c}/q_i\right),
\end{equation}
where $g_{i,c}$ is the shortfall of specification $i$ at corner $c$ and $q_i$ its scale (a log ratio, 10\,dB or $30^\circ$).
Designs are ranked by (infeasible, $V$, tt power, $-$margin, area).
The agent sees each failing specification at its worst corner and flagged devices.

\subsection{Phase-Explore: Agent with Elite Pool}
A tool-calling LLM agent~\cite{react,deepseek} improves verified designs instead of restarting: it edits the netlist, evaluates it, and queries the gm/Id tables.
Each run ends with a lessons-learned summary for later prompts.
An elite pool keeps the $K=3$ best feasible designs.
A new run starts from elite $j$ with probability $p_j\propto\exp(-(r_j-r^\ast)/\sigma_r)$, where $r_j$ is its log-power, $r^\ast$ the best and $\sigma_r$ the spread.
Most runs start from an elite, because local rewrites of a feasible design pay off far more often than fresh starts: after two fresh runs, run $k$ of $N$ starts fresh with probability $\max(0.05,\,0.15(1-k/N))$.
In development, 15.5\% of the agent's evaluation requests repeated the source text of an earlier design exactly. An exact cache therefore returns the stored result of an identical source instead of simulating it again.

\subsection{Phase-Localize: Source-Local Sizing Intent}
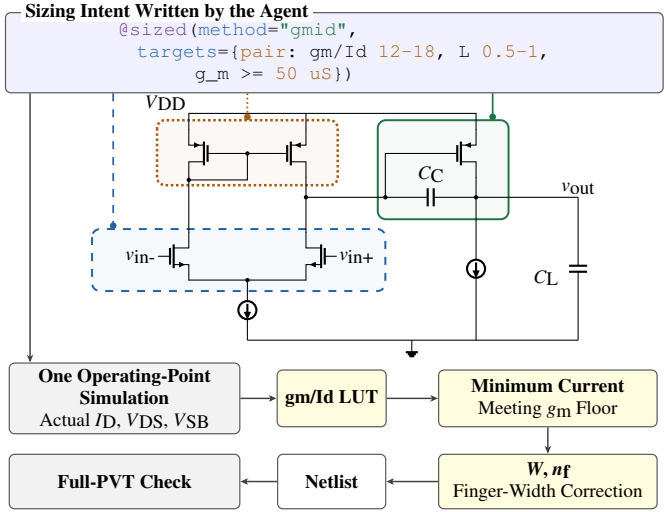
\begin{figure}[t]
\centering
\begin{tikzpicture}[
  x=1cm,y=1cm,font=\fontsize{7.2}{8.3}\selectfont,
  >={Stealth[length=3pt,width=2.5pt]},
  fbox/.style={draw=black!65,line width=.45pt,rounded corners=2pt,
    align=center,inner xsep=3pt,inner ysep=2pt,fill=white},
  flabel/.style={inner sep=1pt,font=\fontsize{7.2}{8.3}\selectfont},
  flow/.style={->,draw=black!75,line width=.55pt},
  link/.style={line width=.65pt,sharp corners},
  pair/.style={draw=f2Blue,dashed,line width=.75pt,rounded corners=3pt},
  mirror/.style={draw=f2Orange,densely dotted,line width=.95pt,rounded corners=3pt},
  stage/.style={draw=f2Green,line width=.75pt,rounded corners=3pt}
]
\definecolor{f2Blue}{HTML}{2462A6}
\definecolor{f2Orange}{HTML}{A85B14}
\definecolor{f2Green}{HTML}{227448}
\definecolor{f2Purple}{HTML}{753D91}
\def\fsub#1#2{\textit{#1}\raisebox{-.30ex}{\fontsize{7.2}{7.2}\selectfont#2}}
\path[use as bounding box] (0,-5.56) rectangle (8.85,1.24);

\node[fbox,fill=blue!6,minimum width=8.70cm,minimum height=1.06cm,
  anchor=south,font=\ttfamily\fontsize{7.2}{8.3}\selectfont]
  (intent) at (4.425,.015) {\begin{tabular}{@{}l@{}}
  \textcolor{f2Purple}{@sized}(\textcolor{f2Blue}{method}=\textcolor{f2Green}{"gmid"},\\
  \quad\textcolor{f2Blue}{targets}=\{\textcolor{f2Orange}{pair}: gm/Id \textcolor{f2Orange}{12-18}, L \textcolor{f2Orange}{0.5-1},\\
  \qquad\qquad g\_m >= \textcolor{f2Orange}{50 uS}\})
  \end{tabular}};
\node[fill=white,inner xsep=1.5pt,inner ysep=.5pt,anchor=west,font=\fontsize{7.2}{8.3}\selectfont\bfseries]
  at ([xshift=6pt]intent.north west) {Sizing Intent Written by the Agent};

\ctikzset{bipoles/length=.53cm,transistors/scale=1.08,sources/scale=.75,capacitors/scale=.75}
\ctikzset{tripoles/mos style/arrows,tripoles/pmos style/nocircle}
\node[pmos,xscale=-1] (P1) at (2.50,-.78) {};
\node[pmos] (P2) at (4.05,-.78) {};
\node[pmos] (P3) at (6.30,-.78) {};
\node[nmos] (N1) at (2.50,-2.14) {};
\node[nmos,xscale=-1] (N2) at (4.05,-2.14) {};
\draw (P1.S) -- (2.50,-.25) -- (6.30,-.25) -- (P3.S);
\draw (P2.S) -- (4.05,-.25);
\node[flabel,anchor=south east] at (2.50,-.25) {\fsub{V}{DD}};
\draw (P1.G) -- (P2.G);
\draw (3.275,-.78) node[circ]{} -- (3.275,-1.13) -- (2.50,-1.13) node[circ]{};
\draw (P1.D) -- (N1.D);
\draw (P2.D) -- (N2.D);
\draw (4.05,-1.36) node[circ]{} -- (5.10,-1.36) node[circ]{}
  -- (5.10,-.78) -- (P3.G);
\draw (5.10,-1.36) to[C] (6.30,-1.36);
\node[flabel,anchor=south] at (5.70,-1.20) {\fsub{C}{C}};
\draw (P3.D) -- (6.30,-1.36) node[circ]{} -- (7.65,-1.36);
\node[flabel,anchor=south] at (7.65,-1.36) {\fsub{v}{out}};
\draw (N1.S) -- (N2.S);
\draw (3.275,0 |- N1.S) node[circ]{} to[I] (3.275,-3.25);
\draw (6.30,-1.36) to[I] (6.30,-3.25);
\draw (7.65,-1.36) to[C] (7.65,-3.25);
\node[flabel,anchor=east] at (7.43,-2.42) {\fsub{C}{L}};
\draw (3.275,-3.25) -- (7.65,-3.25);
\draw (5.45,-3.25) node[ground]{};
\node[flabel,anchor=east] at (N1.G) {\fsub{v}{in-}};
\node[flabel,anchor=west] at (N2.G) {\fsub{v}{in+}};
\begin{scope}[on background layer]
  \draw[mirror,fill=f2Orange!4] (2.08,-.34) rectangle (4.47,-1.20);
  \draw[stage,fill=f2Green!4] (5.00,-.34) rectangle (6.73,-1.65);
  \draw[pair,fill=f2Blue!4] (1.22,-1.78) rectangle (5.10,-2.60);
\end{scope}

\draw[link,f2Blue,dashed,-{Circle[length=2.5pt]}]
  (1.50,.015) -- (1.50,-1.78);
\draw[link,f2Orange,densely dotted,-{Circle[length=2.5pt]}]
  (3.275,.015) -- (3.275,-.34);
\draw[link,f2Green,-{Circle[length=2.5pt]}]
  (6.52,.015) -- (6.52,-.34);

\node[fbox,fill=gray!10,minimum width=3.05cm,minimum height=.75cm]
  (op) at (1.65,-4.02) {\textbf{One Operating-Point}\\\textbf{Simulation}\\
    Actual \fsub{I}{D}, \fsub{V}{DS}, \fsub{V}{SB}};
\node[fbox,fill=yellow!15,minimum width=1.43cm,minimum height=.75cm]
  (lut) at (4.39,-4.02) {\textbf{gm/Id LUT}};
\node[fbox,fill=yellow!15,minimum width=2.89cm,minimum height=.75cm]
  (current) at (7.25,-4.02) {\textbf{Minimum Current}\\Meeting \fsub{g}{m} Floor};
\node[fbox,fill=yellow!15,minimum width=2.89cm,minimum height=.72cm]
  (sizes) at (7.25,-5.12) {\textbf{\textit{W}, \fsub{n}{f}}\\Finger-Width Correction};
\node[fbox,minimum width=1.43cm,minimum height=.72cm]
  (netlist) at (4.39,-5.12) {\textbf{Netlist}};
\node[fbox,fill=gray!10,minimum width=3.05cm,minimum height=.72cm]
  (check) at (1.65,-5.12) {\textbf{Full-PVT Check}};
\draw[flow] (.40,0 |- intent.south) -- (.40,0 |- op.north);
\draw[flow] (op) -- (lut);
\draw[flow] (lut) -- (current);
\draw[flow] (current) -- (sizes);
\draw[flow] (sizes) -- (netlist);
\draw[flow] (netlist) -- (check);
\end{tikzpicture}%

\caption{Phase-Localize on the two-stage OTA: one operating-point simulation measures the actual bias of each \texttt{@sized} block, the gm/Id table gives the minimum current that meets the $g_m$ floor, and the resulting $W$ and $n_f$ are written back before full-PVT verification.}
\label{fig:intent}
\label{fig:concept}
\end{figure}

Instead of tuning sizes by simulation, the agent tags a netlist block with \texttt{@sized} and states its goal as an intent tuple
\begin{equation}
\mathcal{I}=\big(\mathcal{B},\ \min I_{\mathcal{B}},\ [\gamma_{\min},\gamma_{\max}],\ [L_{\min},L_{\max}],\ g_{m,\min}\big),
\end{equation}
where $\mathcal{B}$ is a set of devices, $I_{\mathcal{B}}$ their current and $\gamma$ the gm/Id ratio (Fig.~\ref{fig:intent}).

\textbf{Analytic solver.}
One operating-point simulation of the whole circuit gives each targeted device its \emph{actual} $V_{DS}$ and $V_{SB}$, so the tables are read at the bias the block really sees.
For a stated gm/Id range $[\gamma_{\min},\gamma_{\max}]$, a length range and a floor $g_{m,\min}$, the minimum current meeting the floor is
\begin{equation}
\gamma=\gamma_{\max},\qquad I_D=\max\!\left(I_{\min},\,g_{m,\min}/\gamma\right),
\end{equation}
with $L$ the characterized length nearest the current one.
The total width follows from the tabulated current density, $W_\mathrm{tot}=I_D/J(\gamma,L,V_{DS},V_{SB},W_f)$, iterated to a fixed point because the density of a binned PDK depends on the finger width $W_f=W_\mathrm{tot}/n_f$ for $n_f$ fingers.
Matched devices receive the most conservative common geometry.
The solver then re-measures the bias, using at most three operating-point simulations per block.
If no legal geometry exists, it reports the limiting constraint and the parameter to relax; the evaluator remains the only authority on feasibility.

\textbf{Why analytic.}
Every analog candidate needs multi-corner simulation, so a simulation-driven local optimizer competes with the agent for the same budget (Sec.~\ref{sec:budget}); precomputed device characterization removes that competition.

\textbf{Stage.}
Phase-Localize restarts agent runs from the three best Phase-Explore designs; every evaluated design must call an \texttt{@sized} block placed and targeted by the agent.

\subsection{Phase-Restructure and Phase-Refine}
Phase-Restructure recognizes standard structures in the agent's netlist and swaps each one for another member of its family, for example a wide-swing mirror or Miller-$C$ with a nulling resistor.
Phase-Refine refines the best design with CMA-ES~\cite{cmaes} over matched device groups, within a factor of 4 of its sizes, with restarts on stagnation.
The same engine run from the initial design is the CMA-ES baseline.

\textbf{Sign-off.}
A design is signed off when it passes full PVT, is saturated at all corners with at least 50\,mV margin, reaches a typical-corner mismatch~\cite{pelgrom} Monte Carlo yield of at least 95\% ($N=500$), and independent reference scripts reproduce its measurements.

\begin{table*}[t]
\centering
\caption{\normalfont\fontsize{8}{9.3}\selectfont Best full-PVT-feasible TT power of each method at an equal budget of 16k simulations (a) and of AnalogScout's phases (b).}
\label{tab:ablation}
\footnotesize\setlength{\tabcolsep}{1.3pt}
\begin{tabular*}{\textwidth}{@{\extracolsep{\fill}}p{0.29\textwidth}rrrrrr@{}}
\toprule
 & \multicolumn{2}{c}{\textbf{Miller 2-Stage}} & \multicolumn{2}{c}{\textbf{Folded Cascode}} & \multicolumn{2}{c}{\textbf{3-Stage NMC}} \\
\cmidrule(lr){2-3}\cmidrule(lr){4-5}\cmidrule(l){6-7}
Configuration & \shortstack{Median $P$\\(\textmu W)} & \shortstack{Mean $\pm$ sample std\\(\textmu W)} & \shortstack{Median $P$\\(\textmu W)} & \shortstack{Mean $\pm$ sample std\\(\textmu W)} & \shortstack{Median $P$\\(\textmu W)} & \shortstack{Mean $\pm$ sample std\\(\textmu W)} \\
\midrule
\multicolumn{7}{@{}l}{\textit{(a) Same-budget comparison}} \\
Start & 451 & 451 & 414 & 414 & 2{,}809 & 2{,}809 \\
CMA-ES@16k & 124 & 131\,$\pm$\,16 & 163 & 164\,$\pm$\,10 & 1{,}561 & 1{,}554\,$\pm$\,164 \\
BO@16k & 142 & 140\,$\pm$\,15 & 172 & 172\,$\pm$\,4 & 1{,}955 & 1{,}965\,$\pm$\,103 \\
LAS-Style & 99.3 & 98.8\,$\pm$\,6.0 & 115 & 118\,$\pm$\,7 & -- & -- \\
ACP-Style & 145 & 147\,$\pm$\,6 & 168 & 169\,$\pm$\,7 & -- & -- \\
CMA-ES@32k (2x budget) & 107 & 109\,$\pm$\,6 & -- & -- & 1{,}549 & 1{,}559\,$\pm$\,53 \\
AnalogScout (16k) & \textbf{61.4} & \textbf{65.6\,$\pm$\,14.4} & \textbf{98.1} & \textbf{99.9\,$\pm$\,3.4} & \textbf{944} & \textbf{968\,$\pm$\,111} \\
\midrule
\multicolumn{7}{@{}l}{\textit{(b) Ablation}} \\
Phase-Explore w/o elite pool (8k) & 98.0 & 102\,$\pm$\,8 & 125 & 124\,$\pm$\,4 & -- & -- \\
Phase-Explore (8k) & 78.0 & 77.5\,$\pm$\,8.0 & 105 & 106\,$\pm$\,5 & 1{,}079 & 1{,}054\,$\pm$\,111 \\
+ Phase-Localize (12k) & 62.9 & 66.7\,$\pm$\,13.9 & 98.1 & 100\,$\pm$\,3 & 953 & 976\,$\pm$\,116 \\
Phase-Explore + restart (12k) & 64.6 & 68.4\,$\pm$\,12.8 & 96.9 & 99.3\,$\pm$\,7.0 & 862 & 876\,$\pm$\,164 \\
+ Phase-Restructure (14k) & 62.9 & 66.5\,$\pm$\,14.1 & 98.1 & 100\,$\pm$\,3 & 953 & 976\,$\pm$\,116 \\
+ Phase-Refine (16k) & 61.4 & 65.6\,$\pm$\,14.4 & 98.1 & 99.9\,$\pm$\,3.4 & 944 & 968\,$\pm$\,111 \\
\bottomrule
\end{tabular*}
\par\vspace{2pt}\noindent\parbox{\textwidth}{\raggedright Every reported design meets all specifications of Fig. 4 at all nine PVT corners in the fixed evaluator; $P$: TT power; $n = 5$ searches per method (BO, LAS-Style, ACP-Style, CMA-ES@32k: $n = 3$; Start: one hand design). The NMC CMA-ES@32k entry was added after H1--H3 were fixed and is descriptive only (seeds 4--6; all outcomes reported).}
\end{table*}

\section{Experiments}\label{sec:exp}
\subsection{Setup}
\textbf{Benchmarks.}
We size three single-ended OTAs of increasing complexity in SKY130~\cite{sky130} at $V_{DD}=1.8$\,V, starting from gm/Id hand designs (Fig.~\ref{fig:benchmarks}); the Miller OTA is repeated in GF180MCU~\cite{gf180} at 3.3\,V as a test-only PDK.

\begin{figure}[t]
\centering
\begin{tikzpicture}[x=1cm,y=1cm,font=\fontsize{7.2}{7.3}\selectfont,
  every node/.style={inner sep=0pt},line width=.45pt,
  cell/.style={text height=4.9pt,text depth=1.56pt}]
\def\fsub#1#2{\textit{#1}\raisebox{-.25ex}{\fontsize{7.2}{7.2}\selectfont#2}}
\def\bmath#1{{\fontsize{8}{8}\selectfont\ensuremath{#1}}}
\path[use as bounding box] (0,-3.10) rectangle (8.85,.94);
\node[cell,anchor=west] at (.02,0.82) {\textbf{Specification}};
\node[cell] at (3.15,0.82) {\textbf{Miller 2-Stage}};
\draw (2.322,0.490) -- (2.462,0.490);
\draw (2.322,0.390) -- (2.462,0.390);
\draw (2.462,0.320) -- (2.462,0.560) -- (2.722,0.440) -- cycle;
\draw (2.722,0.440) -- (2.982,0.440);
\draw (2.982,0.320) -- (2.982,0.560) -- (3.242,0.440) -- cycle;
\draw (3.242,0.440) -- (3.442,0.440) -- (3.442,0.375);
\draw (3.372,0.375) -- (3.512,0.375);
\draw (3.372,0.325) -- (3.512,0.325);
\draw (3.442,0.325) -- (3.442,0.255);
\draw (3.392,0.255) -- (3.492,0.255) -- (3.442,0.195) -- cycle;
\node[cell,anchor=west] at (3.542,0.350) {2\,pF};
\draw (2.822,0.440) -- (2.822,0.620) -- (3.107,0.620);
\draw (3.157,0.620) -- (3.442,0.620) -- (3.442,0.440);
\draw (3.107,0.550) -- (3.107,0.690);
\draw (3.157,0.550) -- (3.157,0.690);
\fill (2.822,0.440) circle[radius=.022];
\fill (3.442,0.440) circle[radius=.022];
\node[cell] at (5.42,0.82) {\textbf{Folded Cascode}};
\draw (4.852,0.490) -- (4.992,0.490);
\draw (4.852,0.390) -- (4.992,0.390);
\draw (4.992,0.320) -- (4.992,0.560) -- (5.252,0.440) -- cycle;
\draw (5.252,0.440) -- (5.452,0.440) -- (5.452,0.375);
\draw (5.382,0.375) -- (5.522,0.375);
\draw (5.382,0.325) -- (5.522,0.325);
\draw (5.452,0.325) -- (5.452,0.255);
\draw (5.402,0.255) -- (5.502,0.255) -- (5.452,0.195) -- cycle;
\node[cell,anchor=west] at (5.552,0.350) {2\,pF};
\node[cell] at (7.69,0.82) {\textbf{3-Stage NMC}};
\draw (6.599,0.490) -- (6.739,0.490);
\draw (6.599,0.390) -- (6.739,0.390);
\draw (6.739,0.320) -- (6.739,0.560) -- (6.998,0.440) -- cycle;
\draw (6.998,0.440) -- (7.178,0.440);
\draw (7.178,0.320) -- (7.178,0.560) -- (7.438,0.440) -- cycle;
\draw (7.438,0.440) -- (7.618,0.440);
\draw (7.618,0.320) -- (7.618,0.560) -- (7.878,0.440) -- cycle;
\draw (7.878,0.440) -- (8.119,0.440) -- (8.119,0.375);
\draw (8.049,0.375) -- (8.189,0.375);
\draw (8.049,0.325) -- (8.189,0.325);
\draw (8.119,0.325) -- (8.119,0.255);
\draw (8.069,0.255) -- (8.169,0.255) -- (8.119,0.195) -- cycle;
\node[cell,anchor=west] at (8.219,0.350) {50\,pF};
\draw (7.088,0.440) -- (7.088,0.620) -- (7.293,0.620);
\draw (7.344,0.620) -- (8.119,0.620) -- (8.119,0.440);
\draw (7.293,0.550) -- (7.293,0.690);
\draw (7.344,0.550) -- (7.344,0.690);
\draw (7.528,0.440) -- (7.528,0.260) -- (7.718,0.260);
\draw (7.768,0.260) -- (7.958,0.260) -- (7.958,0.440);
\draw (7.718,0.190) -- (7.718,0.330);
\draw (7.768,0.190) -- (7.768,0.330);
\fill (7.088,0.440) circle[radius=.022];
\fill (7.528,0.440) circle[radius=.022];
\fill (7.958,0.440) circle[radius=.022];
\fill (8.119,0.440) circle[radius=.022];
\draw[line width=.6pt] (0,0.13799999999999976) -- (8.85,0.13799999999999976);
\node[cell,anchor=west] at (.02,-0.022) {\textit{Minimum}};
\node[cell,anchor=west] at (.16,-0.266) {Gain (dB)};
\node[cell] at (3.15,-0.266) {70};
\node[cell] at (5.42,-0.266) {58};
\node[cell] at (7.69,-0.266) {115};
\node[cell,anchor=west] at (.16,-0.510) {GBW (MHz)};
\node[cell] at (3.15,-0.510) {10};
\node[cell] at (5.42,-0.510) {20};
\node[cell] at (7.69,-0.510) {5};
\node[cell,anchor=west] at (.16,-0.754) {PM (deg)};
\node[cell] at (3.15,-0.754) {60};
\node[cell] at (5.42,-0.754) {60};
\node[cell] at (7.69,-0.754) {70};
\node[cell,anchor=west] at (.16,-0.998) {SR (V/\bmath{\mu}s)};
\node[cell] at (3.15,-0.998) {10};
\node[cell] at (5.42,-0.998) {10};
\node[cell] at (7.69,-0.998) {4.5};
\node[cell,anchor=west] at (.16,-1.242) {CMRR (dB)};
\node[cell] at (3.15,-1.242) {60};
\node[cell] at (5.42,-1.242) {60};
\node[cell] at (7.69,-1.242) {50};
\node[cell,anchor=west] at (.16,-1.486) {PSRR (dB)};
\node[cell] at (3.15,-1.486) {55};
\node[cell] at (5.42,-1.486) {55};
\node[cell] at (7.69,-1.486) {40};
\node[cell,anchor=west] at (.16,-1.730) {Swing (V)};
\node[cell] at (3.15,-1.730) {1};
\node[cell] at (5.42,-1.730) {0.9};
\node[cell] at (7.69,-1.730) {1.3};
\node[cell,anchor=west] at (.02,-1.974) {\textit{Maximum}};
\node[cell,anchor=west] at (.16,-2.218) {Settling (ns)};
\node[cell] at (3.15,-2.218) {200};
\node[cell] at (5.42,-2.218) {200};
\node[cell] at (7.69,-2.218) {170};
\node[cell,anchor=west] at (.16,-2.462) {Noise (\bmath{\mu}V rms)};
\node[cell] at (3.15,-2.462) {65};
\node[cell] at (5.42,-2.462) {85};
\node[cell] at (7.69,-2.462) {26};
\node[cell,anchor=west] at (.16,-2.706) {Area (\bmath{\mu}m\raisebox{.45ex}{\fontsize{7.2}{7.2}\selectfont 2})};
\node[cell] at (3.15,-2.706) {5000};
\node[cell] at (5.42,-2.706) {5000};
\node[cell] at (7.69,-2.706) {18000};
\node[cell,anchor=west] at (.16,-2.950) {Power (\bmath{\mu}W)};
\node[cell] at (3.15,-2.950) {600};
\node[cell] at (5.42,-2.950) {550};
\node[cell] at (7.69,-2.950) {3300};
\draw[line width=.6pt] (0,-3.10) -- (8.85,-3.10);
\end{tikzpicture}%

\par\vspace{-4pt}
\caption{Benchmark OTAs and their specifications; every specification must hold at all nine PVT corners, TT power is minimized, and the GF180 variants scale voltage specifications by 3.3/1.8.}
\label{fig:benchmarks}
\end{figure}

\textbf{Budget and seeds.}
Every method receives 16k simulations, where one simulation is one SPICE run of one testbench at one corner in ngspice~\cite{ngspice}, and only evaluations completed within the budget count.
AnalogScout allots 8k simulations to Phase-Explore (60 agent runs of up to 30 steps on \texttt{deepseek-flash}~\cite{deepseek}), 4k to Phase-Localize, 2k to Phase-Restructure and the remainder to Phase-Refine.
We fixed the method and three hypotheses on development seeds 1--3 and report confirmation seeds 4--8, five per OTA and three for GF180.

\textbf{Baselines.}
Global CMA-ES~\cite{cmaes} and BO~\cite{botorch,ament_logei} size all devices from the starting design with the same bound handling and restarts as Phase-Refine, and CMA-ES is also run with twice the budget.
LAS-style~\cite{llmampsizing} and ACP-style~\cite{analogcoderpro} baselines let the LLM propose a design and hand its sizes to a global optimizer on the same evaluator.
H1 tests AnalogScout against CMA-ES at equal budget, H2 Phase-Localize against an equal-budget restart from the same designs, and H3 Phase-Explore with and without the elite pool; all are one-sided and Holm-corrected at $\alpha=0.05$.

\begin{figure*}[!b]
\centering
\includegraphics[width=\textwidth]{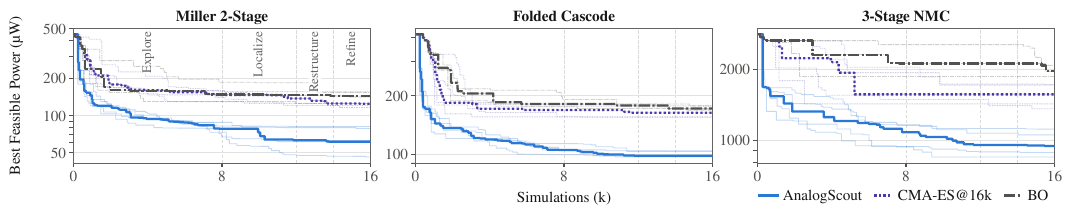}
\par\vspace{-8pt}
\caption{Median best full-PVT-feasible TT power of AnalogScout, CMA-ES and BO versus simulations over five confirmation searches (BO: three); thin lines: individual searches; dashed lines: AnalogScout phase boundaries.}
\label{fig:results}
\label{fig:budget}
\label{fig:convergence}
\end{figure*}

\subsection{Budget Efficiency}
At the same 16k-simulation budget, AnalogScout reaches the lowest power of all methods on every OTA (Table~\ref{tab:ablation}).
It reduces power by 43.3\% relative to CMA-ES (H1, Holm-adjusted $p=\ensuremath{9.16\times10^{-6}}$), and on every OTA each of its five searches beats each CMA-ES search.
On the Miller and NMC OTAs, it still reaches 42.4\% and 39.0\% lower power than CMA-ES with 32k simulations.
The convergence curves in Fig.~\ref{fig:convergence} show that AnalogScout matches the final CMA-ES power after 694--1{,}124 simulations, about 7.0\% of the budget.

\subsection{Where the Budget Goes}\label{sec:budget}
The ablation in Table~\ref{tab:ablation} traces this efficiency to reusing verified designs.
With the elite pool, Phase-Explore reaches 18.1\% lower power than with fresh starts only at 8k simulations (H3, $p=0.00022$).
On development seeds, Phase-Localize spends only 4.3\% of its simulations on sizing, against 85.3--96.3\% for a simulation-based local optimizer, and evaluates about twelve times more designs within the same 4k simulations.
At equal budget, however, it shows no significant power reduction over restarting from the same designs (H2, $p=0.645$), so we claim its low simulation use, not a power gain.
Phase-Restructure leaves the median power unchanged, and Phase-Refine lowers it by at most a further 2.4\%.

\subsection{Generalization and Sign-off}
AnalogScout's designs pass independent all-corner re-measurement about as often as CMA-ES's.
Independent scripts re-measure 156 reference cases from all methods; 46 fail. Per method, 29 of AnalogScout's 111 cases fail, as do 3 of CMA-ES's 12, 2 of BO's 11 and 14 of the LLM baselines' 31 (shared cases count for each).
Forty of the failures only miss the 50\,mV all-corner saturation margin, which the loop checks at the typical corner only; the other 6 exceed a specification by at most 0.5\%.
Among the best designs, which also undergo typical-corner mismatch Monte Carlo ($N=500$), the SKY130 Miller design is signed off, the folded-cascode design fails the re-measurement and reaches a 91.4\% Monte Carlo yield (target 95\%), and the NMC design misses the 170\,ns settling specification by 0.15\% at one corner.
On GF180MCU, AnalogScout reduces Miller OTA power by 59.6\% versus CMA-ES, and its best design is signed off with a 99.8\% Monte Carlo yield.

\textbf{Limitations.}
Phase-Localize shows no significant power reduction over an equal-budget restart, Phase-Restructure adds little, and the GF180 test covers only the Miller OTA.

\section{Conclusion}\label{sec:concl}
AnalogScout sizes analog circuits across PVT corners budget-efficiently by refining verified designs and resolving local sizing intent with almost no simulations.
Future work will resize blocks analytically after structural changes and check saturation at all corners inside the loop; the code and open-PDK data will be released.

\section*{Acknowledgment}
AI coding assistants (ChatGPT) helped implement the software; the authors reviewed it and take full responsibility for its content.


\clearpage
\begin{thebibliography}{10}
\providecommand{\url}[1]{#1}
\csname url@samestyle\endcsname
\providecommand{\newblock}{\relax}
\providecommand{\bibinfo}[2]{#2}
\providecommand{\BIBentrySTDinterwordspacing}{\spaceskip=0pt\relax}
\providecommand{\BIBentryALTinterwordstretchfactor}{4}
\providecommand{\BIBentryALTinterwordspacing}{\spaceskip=\fontdimen2\font plus
\BIBentryALTinterwordstretchfactor\fontdimen3\font minus
  \fontdimen4\font\relax}
\providecommand{\BIBforeignlanguage}[2]{{%
\expandafter\ifx\csname l@#1\endcsname\relax
\typeout{** WARNING: IEEEtran.bst: No hyphenation pattern has been}%
\typeout{** loaded for the language `#1'. Using the pattern for}%
\typeout{** the default language instead.}%
\else
\language=\csname l@#1\endcsname
\fi
#2}}
\providecommand{\BIBdecl}{\relax}
\BIBdecl

\bibitem{gielen_cad}
G.~G.~E. Gielen and R.~A. Rutenbar, ``Computer-aided design of analog and
  mixed-signal integrated circuits,'' \emph{Proc. IEEE}, vol.~88, no.~12, pp.
  1825--1854, Dec. 2000.

\bibitem{wang_bo}
W.~Lyu \emph{et~al.}, ``An efficient {Bayesian} optimization approach for
  automated optimization of analog circuits,'' \emph{IEEE Trans. Circuits Syst.
  I, Reg. Papers}, vol.~65, no.~6, pp. 1954--1967, Jun. 2018.

\bibitem{lyu_mace}
W.~Lyu, F.~Yang, C.~Yan, D.~Zhou, and X.~Zeng, ``Batch {Bayesian} optimization
  via multi-objective acquisition ensemble for automated analog circuit
  design,'' in \emph{Proc. 35th Int. Conf. Mach. Learn. (ICML)}, ser. Proc.
  Mach. Learn. Res., vol.~80, Jul. 2018, pp. 3306--3314.

\bibitem{guo_bnnbyo}
N.~Guo, F.~Peng, J.~Shi, F.~Yang, J.~Tao, and X.~Zeng, ``Yield optimization for
  analog circuits over multiple corners via {Bayesian} neural networks:
  Enhancing circuit reliability under environmental variation,'' \emph{ACM
  Trans. Des. Autom. Electron. Syst.}, vol.~29, no.~1, pp. 12:1--12:17, 2024.

\bibitem{liu_gaspad}
B.~Liu, D.~Zhao, P.~Reynaert, and G.~G.~E. Gielen, ``{GASPAD}: A general and
  efficient mm-wave integrated circuit synthesis method based on surrogate
  model assisted evolutionary algorithm,'' \emph{IEEE Trans. Comput.-Aided
  Design Integr. Circuits Syst.}, vol.~33, no.~2, pp. 169--182, Feb. 2014.

\bibitem{autockt}
K.~Settaluri, A.~Haj-Ali, Q.~Huang, K.~Hakhamaneshi, and B.~Nikolic,
  ``{AutoCkt}: Deep reinforcement learning of analog circuit designs,'' in
  \emph{Proc. Design, Autom. Test Eur. Conf. Exhib. (DATE)}, Mar. 2020, pp.
  490--495.

\bibitem{gcnrl}
H.~Wang \emph{et~al.}, ``{GCN-RL} circuit designer: Transferable transistor
  sizing with graph neural networks and reinforcement learning,'' in
  \emph{Proc. 57th ACM/IEEE Design Autom. Conf. (DAC)}, Jul. 2020, pp. 1--6.

\bibitem{li_pvttransfer}
J.~Li, Y.~Zeng, H.~Zhi, J.~Yang, W.~Shan, Y.~Li, and Y.~Li, ``Knowledge
  transfer framework for {PVT} robustness in analog integrated circuits,''
  \emph{IEEE Trans. Circuits Syst. I, Reg. Papers}, vol.~71, no.~5, pp.
  2017--2030, 2024.

\bibitem{shi_robustanalog}
W.~Shi, H.~Wang, J.~Gu, M.~Liu, D.~Z. Pan, S.~Han, and N.~Sun,
  ``{RobustAnalog}: Fast variation-aware analog circuit design via multi-task
  {RL},'' in \emph{Proc. ACM/IEEE Workshop Mach. Learn. CAD (MLCAD)}, 2022, pp.
  35--41.

\bibitem{analogcoder}
Y.~Lai \emph{et~al.}, ``{AnalogCoder}: Analog circuit design via training-free
  code generation,'' in \emph{Proc. AAAI Conf. Artif. Intell. (AAAI)}, vol.~39,
  no.~1, 2025, pp. 379--387.

\bibitem{artisan}
Z.~Chen \emph{et~al.}, ``Artisan: Automated operational amplifier design via
  domain-specific large language model,'' in \emph{Proc. 61st ACM/IEEE Design
  Autom. Conf. (DAC)}, Jun. 2024, pp. 1--6.

\bibitem{liu_newcas}
C.~Liu, E.~A. Olowe, and D.~Chitnis, ``{LLM}-based {AI} agent for sizing of
  analog and mixed signal circuit,'' in \emph{Proc. IEEE Int. New Circuits
  Syst. Conf. (NEWCAS)}, 2025, arXiv:2504.11497.

\bibitem{analogcoderpro}
Y.~Lai \emph{et~al.}, ``{AnalogCoder-Pro}: Unifying analog circuit generation
  and optimization via multi-modal {LLMs},'' \emph{IEEE Trans. Comput.-Aided
  Design Integr. Circuits Syst.}, 2026, early access, doi:
  10.1109/TCAD.2026.3673493.

\bibitem{adollm}
Y.~Yin, Y.~Wang, B.~Xu, and P.~Li, ``{ADO-LLM}: Analog design {Bayesian}
  optimization with in-context learning of large language models,'' in
  \emph{Proc. 43rd IEEE/ACM Int. Conf. Comput.-Aided Des. (ICCAD)}, Oct. 2024,
  pp. 1--9.

\bibitem{eesizer}
C.~Liu and D.~Chitnis, ``{EEsizer}: {LLM}-based {AI} agent for sizing of analog
  and mixed signal circuit,'' \emph{IEEE Trans. Circuits Syst. I, Reg. Papers},
  vol.~73, no.~6, pp. 3825--3834, Jun. 2026.

\bibitem{llmampsizing}
\BIBentryALTinterwordspacing
J.~Duan \emph{et~al.}, ``{LLM}-assisted analog amplifier sizing,'' GitHub
  repository, 2026, accessed: Oct. 22, 2026. [Online]. Available:
  \url{https://github.com/jiyuanduan001-oss/LLM_Amplifier_Sizing}
\BIBentrySTDinterwordspacing

\bibitem{rtlscout}
F.~Arnold, R.~Amaudruz, D.~Tsaras, R.~Andri, and L.~Cavigelli, ``{RTLScout}:
  Joint agentic code and synthesis optimization for efficient digital
  circuits,'' arXiv:2606.06530, 2026.

\bibitem{huang_croza}
J.~Huang, J.~Shen, F.~Yang, L.~Shang, Z.~Bi, C.~Yan, W.~Hu, D.~Zhou, and
  X.~Zeng, ``Variation-aware analog circuit design via contextual modeling and
  robust optimization,'' \emph{ACM Trans. Des. Autom. Electron. Syst.},
  vol.~31, no.~2, pp. 26:1--26:19, 2026.

\bibitem{kong_pvtsizing}
Z.~Kong, X.~Tang, W.~Shi, Y.~Du, Y.~Lin, and Y.~Wang, ``{PVTSizing}: A
  {TuRBO-RL}-based batch-sampling optimization framework for {PVT}-robust
  analog circuit synthesis,'' in \emph{Proc. ACM/IEEE Design Autom. Conf.
  (DAC)}, 2024, pp. 47:1--47:6.

\bibitem{pascu_smartcorner}
O.~Pascu, C.~Visan, G.~Nicolae, M.~Boldeanu, H.~Cucu, C.~Diaconu, A.~Buzo, and
  G.~Pelz, ``Efficient multi-objective optimization for {PVT} variation-aware
  circuit sizing using surrogate models and smart corner sampling,'' in
  \emph{Proc. IEEE/ACM Int. Symp. Low Power Electron. Design (ISLPED)}, 2023,
  pp. 1--6.

\bibitem{silveira_gmid}
F.~Silveira, D.~Flandre, and P.~G.~A. Jespers, ``A {$g_m/I_D$} based
  methodology for the design of {CMOS} analog circuits and its application to
  the synthesis of a silicon-on-insulator micropower {OTA},'' \emph{IEEE J.
  Solid-State Circuits}, vol.~31, no.~9, pp. 1314--1319, Sep. 1996.

\bibitem{gmid}
P.~G.~A. Jespers and B.~Murmann, \emph{Systematic Design of Analog {CMOS}
  Circuits: Using Pre-Computed Lookup Tables}.\hskip 1em plus 0.5em minus
  0.4em\relax Cambridge, U.K.: Cambridge University Press, 2017.

\bibitem{murmann_iscas24}
B.~Murmann, ``Practical aspects of script-based analog design using precomputed
  lookup tables,'' in \emph{Proc. IEEE Int. Symp. Circuits Syst. (ISCAS)},
  2024.

\bibitem{react}
S.~Yao \emph{et~al.}, ``{ReAct}: Synergizing reasoning and acting in language
  models,'' in \emph{Proc. Int. Conf. Learn. Represent. (ICLR)}, 2023.

\bibitem{deepseek}
{DeepSeek-AI}, ``{DeepSeek-V4}: Towards highly efficient million-token context
  intelligence,'' arXiv:2606.19348, 2026.

\bibitem{cmaes}
N.~Hansen, ``The {CMA} evolution strategy: A tutorial,'' arXiv:1604.00772,
  2016.

\bibitem{pelgrom}
M.~J.~M. Pelgrom, A.~C.~J. Duinmaijer, and A.~P.~G. Welbers, ``Matching
  properties of {MOS} transistors,'' \emph{IEEE J. Solid-State Circuits},
  vol.~24, no.~5, pp. 1433--1439, Oct. 1989.

\bibitem{sky130}
\BIBentryALTinterwordspacing
{SkyWater Technology Foundry and Google}, ``{SkyWater} open source {PDK},''
  GitHub repository, 2020, accessed: Oct. 22, 2026. [Online]. Available:
  \url{https://github.com/google/skywater-pdk}
\BIBentrySTDinterwordspacing

\bibitem{gf180}
\BIBentryALTinterwordspacing
{GlobalFoundries and Google}, ``{GlobalFoundries} {GF180MCU} open source
  {PDK},'' GitHub repository, 2022, accessed: Oct. 22, 2026. [Online].
  Available: \url{https://github.com/google/gf180mcu-pdk}
\BIBentrySTDinterwordspacing

\bibitem{ngspice}
\BIBentryALTinterwordspacing
H.~Vogt, G.~Atkinson, D.~Warning, and P.~Nenzi, \emph{Ngspice User's Manual,
  Version 47}, Aug. 2026. [Online]. Available:
  \url{https://ngspice.sourceforge.io/docs/ngspice-47-manual.pdf}
\BIBentrySTDinterwordspacing

\bibitem{botorch}
M.~Balandat \emph{et~al.}, ``{BoTorch}: A framework for efficient {Monte-Carlo}
  {Bayesian} optimization,'' in \emph{Proc. Adv. Neural Inf. Process. Syst.
  (NeurIPS)}, vol.~33, 2020, pp. 21\,524--21\,538.

\bibitem{ament_logei}
S.~Ament, S.~Daulton, D.~Eriksson, M.~Balandat, and E.~Bakshy, ``Unexpected
  improvements to expected improvement for {Bayesian} optimization,'' in
  \emph{Proc. Adv. Neural Inf. Process. Syst. (NeurIPS)}, vol.~36, 2023, pp.
  20\,577--20\,612.

\end{thebibliography}
\end{document}